Electron-beam evaporated cobalt films on molecular beam epitaxy prepared GaAs(001)

Z. Ding and P.M. Thibado
*Department of Physics, The University of Arkansas, Fayetteville, Arkansas 72701*

C. Awo-Affouda and V.P. LaBella
*School of NanoSciences and NanoEngineering, State University of New York at Albany, Albany, New York 12203*

We have deposited Co films on the GaAs(001) surface by using an e-beam evaporation method. The thicknesses of the Co films are measured by using x-ray reflectivity and Rutherford backscattering. The magnetic properties of the films have been measured using superconducting quantum interference device. The magnetization of the films was found to decrease with increasing film thickness. The slight degradation of magnetic properties is attributed to increasing roughness on the Co surface or the Co/GaAs interface during the Co deposition.

I. INTRODUCTION

Research in the area of spintronic devices made using ferromagnetic/semiconductor heterostructures is of growing interest.[1] For example, there is a current drive to make a spin filter device.[2] It is also believed that spintronic devices will predominate in the electronics industry in the future. Injection of electrons from a magnetic material to a semiconductor is considered to be the most challenging task in spintronics. To achieve high efficiency of spin injection and high spin polarization, it is thought that an atomically abrupt interface is important.[3] Therefore; it is essential to study the electronic and magnetic properties of ferromagnets in ferromagnetic/semiconductor structures.

One approach to developing spintronic devices is to replace the normal metal contacts used on semiconductor device structures with ferromagnetic metal contacts. This can provide a source of spin-polarized current. Ideally, one wants to use a ferromagnetic metal that provides 100% spin-polarized current, shares the same crystal structure, and shares the same lattice constant as the semiconductor. Heusler alloys are a group of ferromagnetic metals that have 100% spin-polarized electrons at the Fermi level.[4] Several share a compatible crystal structure and lattice constant with the III–V semiconductors, such as $Co_2MnAl$.

Another key factor in developing spintronics is the quality of the interface between the ferromagnetic metal and the semiconductor. The interface between the semiconductor and the ferromagnetic material plays a critical role in the efficiency of spin transmission.[5] The structure of $Co_2MnAl$ consists of alternating planes of Co and MnAl. Thus, it may be important to first study Co grown on GaAs(001). In this article, we report on the quality of these films and compare them to bulk GaAs and bulk cobalt. In addition the evolution of the interface and film microstructure during growth and its correlation to magnetic behavior are discussed.

II. EXPERIMENT

Experiments were performed in an ultrahigh vacuum (UHV) multi-chamber facility ($5-8 \times 10^{-11}$ Torr throughout) which contains a solid-source molecular beam epitaxy (MBE) chamber (Riber 32P) with a substrate temperature determination system accurate to +/- 2 °C,[6] and an arsenic cell with an automated valve and controller. The MBE chamber also has an all UHV connection to a surface analysis chamber, which contains a custom integrated scanning tunneling microscope (STM) (Omicron).[7] Commercially available ''epi-ready'', $n+$ (Si doped $10^{18}/cm^3$) GaAs(001) +/- 0.05° substrates were loaded into the MBE system without any chemical cleaning. The surface oxide layer was removed and a 1.5-μm-thick GaAs buffer layer was grown at 580 °C using an $As_4$ to Ga beam equivalent pressure (BEP) ratio of 15 and a growth rate of 1.0 μm/h as determined by reflection high-energy electron diffraction (RHEED) oscillations. After growth, the surface was annealed under an $As_4$ BEP of 1 μTorr for 15 min at 600 °C followed by another at 570 °C under the same conditions. This procedure improves the RHEED pattern and prepares the surface for STM measurement and e-beam evaporation of Co. For STM observation of the surface morphology, the samples were transferred to the STM without breaking UHV, and imaged at room temperature.

After imaging the GaAs(001) surface with the STM, the substrate was transferred to another chamber equipped with a mini e-beam evaporator, without breaking UHV, for the deposition of Co. The experimental set up for the e-beam evaporation is illustrated in Fig. 1. During the evaporation of Co, the GaAs substrate was kept at room temperature and the deposition rate of Co was measured using a quartz crystal monitor. The electron beam is emitted from a hot filament and accelerated by a high voltage to hit the target Co material and heat it. Before the deposition of the Co on the sample, the quartz crystal monitor is placed in the position of the sample to measure the deposition rate of the Co. After the measurement of the deposition rate, the quartz crystal monitor is moved to an adjacent location where a reduced deposition rate can be monitored and scaled to yield the original rate. The sample is then positioned in the same place that the quartz crystal monitor was originally placed, so the deposition rate of the Co is known. By doing this the deposition rate of the Co is believed to be the same as measured by the quartz crystal monitor. By controlling the deposition time, different thicknesses of the Co film can be obtained.

The thicknesses of the Co films were measured by both x-ray reflectivity (XRR) and Rutherford backscattering (RBS) experiments. These also provide a calibration for the calculated quartz crystal monitor measurement of the Co film thickness. The magnetic properties of the Co films on GaAs were measured by a superconducting quantum interference device (SQUID). The magnetization of the Co films is calculated from the SQUID data combined with the RBS data.

III. RESULTS

The *ex situ* XRR and RBS experiments provided the thickness of the Co films as shown in Fig. 2 (RBS data not shown). These results are in good agreement with each other and are consistent with the result from the deposition time and rate obtained by the quartz crystal monitor. In the XRR data, the reflectivity of the Co film oscillates with the incident angle of the x-ray beam. The adjacent oscillation peak spacing is related to the thickness of the film, while the decreasing intensity of oscillations indicates the presence of roughness on the surface and/or the interface. Typically, surface roughness decreases the intensity of the whole curve, while interfacial roughness is primarily responsible for the decay in the amplitude of oscillations. From the STM images we know the starting GaAs surface is extremely flat[8] (in Ref. 8 see STM picture labeled $T_l$) so the roughening is generated by either the cobalt not growing smoothly or is strain induced.

The magnetic properties of the Co films were measured by SQUID. Hysteresis curves are shown in Fig. 3 for the 50 and 560 nm thick Co films acquired at room temperature. The shape of the hysteresis curve near zero field is nearly a rectangle for the Co film with 50 nm thickness. The magnetic moment of the thicker Co film is much higher (almost ten times higher) than that of the thin film. Interestingly, the magnetic moment under higher magnetic field decreases slowly with the applied field for the thin film, while for the thick films, the magnetic moment increases slowly with the applied field.

The magnetization, which is derived from the SQUID and RBS data of the Co films, is shown in Fig. 4. The thicknesses of the three samples are 50, 350, and 560 nm, respectively. The magnetization decreases linearly with the increase of thickness of Co film as shown in the figure. The line that is fit by the least-squares method has been added to the figure. Finally, the magnetization of bulk Co is also shown in the figure near the top as a reference.[9,10]

IV. DISCUSSION

The thicknesses of the Co films were measured *ex situ* by RBS and XRR. The disadvantage of an *ex situ* measurement is that the Co films are exposed to air and this can introduce contamination. By using the quartz crystal monitor, one can determine the thickness of the Co films *in situ* by measuring the deposition time and rate. On the other hand, the XRR and RBS measurements are a direct thickness measurement and can provide a calibration of the thickness. In the XRR data, the oscillations have damping intensities for the peaks, this would indicate the presence of interface roughness between deposited Co film and GaAs substrate as seen in Fig. 2. This may be due to the deposition of Co being carried out at room temperature. The interface roughness may also develop with the increase in film thickness. This is because of the mismatch of lattice constant for Co and GaAs, which produces strain on the interface of Co and GaAs. The strain will cause some defects such as dislocations or stacking faults after the thickness of Co film is grown thicker than a critical value of about 5 nm.[11] This critical value represents the film thickness when the morphology changes from bcc to hcp, while the critical thickness of 50 nm represents the onset of dislocation formation in the film. Our thinnest film is around the critical thickness, while the other two films are seven and ten times this value. As the film grows beyond the critical thickness the dislocation density will increase, and this may play a role in the degradation of the magnetic properties.

By comparison of the hysteresis curves of different samples, one can uncover the magnetic properties of the Co films. Also, one can speculate on the surface roughness of the Co films and the interface roughness between the Co and the GaAs. The interface roughness will affect the crystalline quality of the Co films, which will affect the magnetic properties of Co film in return. From Fig. 4, one can see that the magnetization of the thin film is higher than that of the thick film; this indicates that the thinner film has better crystalline quality and lower surface and/or interfacial roughness. The decrease in magnetization may also indicate an anisotropy in the Co film exists. Practical applications of ferromagnetic materials may require a magnetic thin film to have a single magnetic domain state with a preferential magnetization axis. This feature is typical for a film with good surface ordering. The material needs to be a single crystal and the surface should be well ordered over a long range, which should be achievable with cobalt since it only has a lattice mismatch of 0.2% yielding a critical thickness of about 50 nm before dislocations will form.[12] A film with a disordered surface, by contrast, may alter the magnetic phase on the Co film.[13] In some experiments about magnetic anisotropy of Co films, the experiments and theoretical calculation show that Co films on GaAs(001) and GaAs(110) are in the bcc phase during the early stage of growth.[11,14–16] The bcc phase is believed to originate from imperfections such as overlayer–substrate interaction, substrate atoms segregation, and surface roughness.[11] After the thickness of the Co film is thicker than a critical thickness of about 5 nm, there coexist a polycrystalline hexagonal close packing (hcp) and bcc structures.[12] Some other studies show that after the growth of bcc Co on GaAs(001) in the early stages, there exist two perpendicularly oriented hcp Co domains in the final stages.[17,18] The hcp layer is found to be directly on top of the bcc layer in some other studies.[19,20] For our samples, the thicknesses are much higher than 5 nm, so the crystalline phase of Co films may be hcp on top of bcc.

The growth of Co on the oxide-desorbed GaAs(001) surface is more three-dimensional growth mode similar to that observed for Fe on GaAs(001) because the surface after oxide removal is still rough.[21] For our samples, the starting surface is extremely flat and the starting surface roughness should be very low. Also, the surface reconstruction of the GaAs(001) surface may play a role in the roughening of the deposited Co film. Due to the $2\times4$ reconstruction,[22] the surface might not provide a template as suitable as the (110) surface for driving Co atoms in the metastable bcc phase.[23] In our samples, the Co films are deposited onto the GaAs fresh surface at room temperature and not annealed after the e-beam evaporation, so that the Co/GaAs interface reaction and interdiffusion should be negligible. We speculate that the existence of polycrystalline hcp structure may originate from the low deposition temperature, because at low temperature the Co atoms at GaAs do not have enough energy to reorganize themselves. Thus the surface roughens after the deposition of the first layer of Co atoms and the roughness develops with the proceeding of the deposition process, resulting in the polycrystalline bcc and hcp structures and differently oriented magnetic domains. The existence of different Co domains results in the degradation of magnetization of the Co films compared to the bulk Co material.

## V. CONCLUSION

We have deposited Co films on the GaAs(001) surface by using an e-beam evaporation method. The magnetic properties of the Co films have been measured by SQUID combined with RBS. The magnetization of the Co films is found to decrease with the increase of film thickness. The degradation of the magnetization of the Co film with the increase of film thickness is attributed to the roughness on the Co surface and/or Co/GaAs interface during the Co deposition.

ACKNOWLEDGMENTS

This work is supported by the National Science Foundation Grant No. FRG DMR-0102755 and the State of New York Grant No. NYSTAR-FDP-C020095.

References

[1] G. A. Prinz, Phys. Today 48, 58 (1995).
[2] G. Kirczenow, Phys. Rev. B 63, 054422/1 (2001).
[3] V. P. LaBella, D. W. Bullock, Z. Ding, C. Emery, A. Venkatesan, W. F. Oliver, G. J. Salamo, P. M. Thibado, and M. Mortazavi, Science 292, 1518 (2001).
[4] M. S. Lund, J. W. Dong, J. Lu, X. Y. Dong, C. J. Palmstrom, and C. Leighton, Appl. Phys. Lett. 80, 4798 (2002).
[5] A. T. Hanbicki, B. T. Jonker, G. Itskos, G. Kioseoglou, and A. Petrou, Appl. Phys. Lett. 80, 1240 (2002).
[6] P. M. Thibado, G. J. Salamo, and Y. Baharav, J. Vac. Sci. Technol. B 17, 253 (1999).
[7] J. B. Smathers, D. W. Bullock, Z. Ding, G. J. Salamo, P. M. Thibado, B. Gerace, and W. Wirth, J. Vac. Sci. Technol. B 16, 3112 (1998).
[8] V. P. LaBella, D. W. Bullock, C. Emery, Z. Ding, and P. M. Thibado, Appl. Phys. Lett. 79, 3065 (2001).
[9] C. Kittel, *Introduction to Solid State Physics* (Wiley, New York, 1953).
[10] N. W. Ashcroft and N. D. Mermin, *Solid State Physics* (Saunders College Press, Philadelphia, 1976).
[11] A. Y. Liu and D. J. Singh, J. Appl. Phys. 73, 6189 (1993).
[12] C. J. Palmstrom, C. C. Chang, A. Yu, G. J. Galvin, and J. W. Mayer, J. Appl. Phys. 62, 3755 (1987).
[13] E. J. Escorcia-Aparicio, H. J. Choi, R. K. Kawakami, and Z. Q. Qiu, Phys. Rev. B 58, 93 (1998).
[14] G. A. Prinz, Phys. Rev. Lett. 54, 1051 (1985).
[15] F. Xu, J. J. Joyce, M.W. Ruckman, H.W. Chen, F. Boscherini, D. M. Hill, S. A. Chambers, and J. H. Weaver, Phys. Rev. B 35, 2375 (1987).
[16] Y. U. Idzerda, W. T. Elam, B. T. Jonker, and G. A. Prinz, Phys. Rev. Lett. 62, 2480 (1989).
[17] Y. Z. Wu *et al.*, Phys. Rev. B 57, 11935 (1998).
[18] E. Gu *et al.*, Phys. Rev. B 52, 14704 (1995).
[19] M. A. Mangan, G. Spanos, T. Ambrose, and G. A. Prinz, Appl. Phys. Lett. 75, 346 (1999).
[20] K. G. Nath, F. Maeda, S. Suzuki, and Y. Watanabe, J. Appl. Phys. 90, 1222 (2001).
[21] B. T. Jonker, G. A. Prinz, and Y. U. Idzerda, J. Vac. Sci. Technol. B 9, 2437 (1991).
[22] V. P. LaBella, H. Yang, D. W. Bullock, P. M. Thibado, P. Kratzer, and M. Scheffler, Phys. Rev. Lett. 83, 2989 (1999).
[23] S. J. Blundell, M. Gester, J. A. C. Bland, C. Daboo, E. Gu, M. J. Baird, and A. J. R. Ives, J. Appl. Phys. 73, 5948 (1993).

FIG. 1. Illustration of the e-beam evaporation system. The quartz crystal monitor is used to measure the deposition rate of Co, and can be positioned in the same location as the sample.

FIG. 2. X-ray reflectivity intensity as a function of glancing incident angle for a 50 nm cobalt film grown on GaAs(001).

FIG. 3. Hysteresis loops of Co films obtained by SQUID measurement. The data were acquired at room temperature and the thicknesses of the films are 50 and 560 nm.

FIG. 4. Magnetization of Co films with different thicknesses deposited by e-beam evaporation. The horizontal solid line shown near the top is the magnetization of bulk Co material.

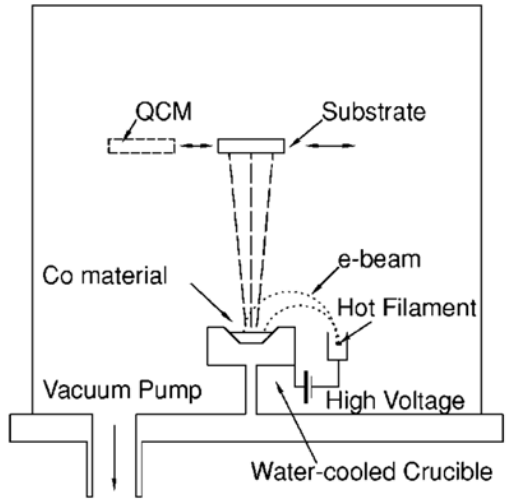

Figure 1.

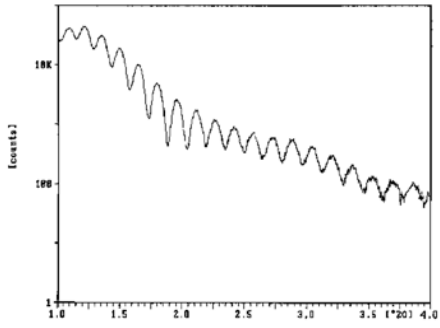

Figure 2.

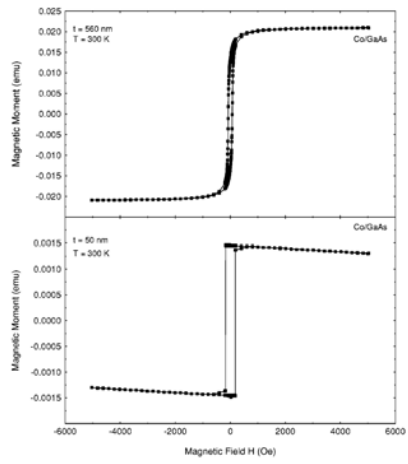

Figure 3.

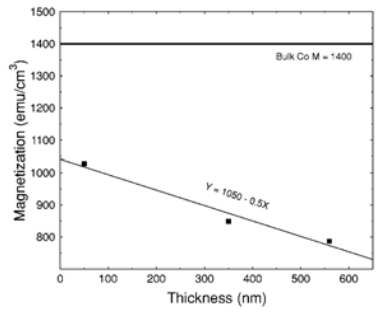

Figure 4.